\documentclass[sigconf]{acmart}

\usepackage{amsmath}
\usepackage[capitalise,noabbrev]{cleveref}
\usepackage{subfigure}
\usepackage{siunitx}  
\graphicspath{{figures/}}

\copyrightyear{2023}
\acmYear{2023}
\acmConference[WNS3 2023]{Proceedings of the WNS3 2023}{June 28--29, 2022}{Washington DC, USA}
\acmBooktitle{Proceedings of the WNS3 2023 (WNS3 2023), June 28--29, 2022, Washington DC, USA}

\begin{document}

\title{Position-Based Machine Learning Propagation Loss Model Enabling Fast Digital Twins of Wireless Networks in ns-3}

\author{Eduardo Nuno Almeida}
\affiliation{
    \institution{INESC TEC and Faculdade de Engenharia, Universidade do Porto}
    \city{Porto}
    \country{Portugal}
}
\email{eduardo.n.almeida@inesctec.pt}

\author{Helder Fontes}
\affiliation{
    \institution{INESC TEC and Faculdade de Engenharia, Universidade do Porto}
    \city{Porto}
    \country{Portugal}
}
\email{helder.m.fontes@inesctec.pt}

\author{Rui Campos}
\affiliation{
    \institution{INESC TEC and Faculdade de Engenharia, Universidade do Porto}
    \city{Porto}
    \country{Portugal}
}
\email{rui.l.campos@inesctec.pt}

\author{Manuel Ricardo}
\affiliation{
    \institution{INESC TEC and Faculdade de Engenharia, Universidade do Porto}
    \city{Porto}
    \country{Portugal}
}
\email{mricardo@inesctec.pt}

\begin{abstract}

Digital twins have been emerging as a hybrid approach that combines the benefits of simulators with the realism of experimental testbeds. The accurate and repeatable set-ups replicating the dynamic conditions of physical environments, enable digital twins of wireless networks to be used to evaluate the performance of next-generation networks. In this paper, we propose the Position-based Machine Learning Propagation Loss Model (P-MLPL), enabling the creation of fast and more precise digital twins of wireless networks in ns-3. Based on network traces collected in an experimental testbed, the P-MLPL model estimates the propagation loss suffered by packets exchanged between a transmitter and a receiver, considering the absolute node's positions and the traffic direction. The P-MLPL model is validated with a test suite. The results show that the P-MLPL model can predict the propagation loss with a median error of 2.5 dB, which corresponds to 0.5x the error of existing models in ns-3. Moreover, ns-3 simulations with the P-MLPL model estimated the throughput with an error up to 2.5 Mbit/s, when compared to the real values measured in the testbed.

\end{abstract}



\maketitle


\section{Introduction} \label{Introduction-Section}


The development of next-generation networks requires the evaluation of their performance in realistic conditions. Experimental testbeds provide real results of the solution's performance, at the expense of cost and complexity of the set-up, as well as the testbed's limited availability. Network simulators, such as ns-3 \cite{henderson2008network}, enable the development of repeatable and reproducible set-ups, which are relatively simple to create. However, the available networking models are generic and do not capture the specific characteristics of a given physical environment, especially in extreme scenarios with dynamic and unknown environment conditions.

In recent years, digital twins have been emerging as a hybrid approach that combines the benefits of simulators with the realism of experimental testbeds \cite{khan2022digital, wu2021digital}. Digital twins are composed of digital models that replicate the behavior of physical systems and the dynamic conditions of experimental environments. As such, they can be used to validate networking solutions and evaluate their performance in simulated environments that realistically replicate the dynamic conditions that characterize the physical environments. On the other hand, digital twins can be used in applications that require large quantities of data and interaction with the environment. One example are Reinforcement Learning algorithms, where agents are trained by applying actions on the environment and collecting observations that result from their actions, which would be very difficult to implement in experimental testbeds \cite{li2022when, almeida2022traffic}. Moreover, digital twins can be used to simulate scaled-up versions of experimental testbeds, with larger and more complex network topologies, that would entail technical difficulties to implement in the physical environment.


One of the key components of wireless networks' digital twins is the wireless channel model. Machine Learning (ML) techniques can be used to create accurate custom models of the wireless channel in a given physical environment, enabling the estimation of the channel quality \cite{cerar2021machine, almeida2019machine}. The ML models are trained with datasets of experimental received power or Signal-to-Noise Ratio (SNR), in order to learn and replicate the dynamic conditions of the environment without requiring predetermined models.

In \cite{almeida2022machine}, we presented the ML-based Propagation Loss (MLPL) model for ns-3. Unlike the trace-based simulation approach \cite{fontes2017trace, cruz2020reproduction}, the MLPL model reproduces, in simulation, the experimental conditions measured in the physical environment for any network topology, node mobility and simulation duration. Using ML models trained with a dataset of network traces collected in an experimental testbed, the MLPL model estimates the radio propagation loss between two nodes for a given distance and time instant. This value is then used by ns-3 to calculate the received power at the destination node, considering the transmission power and the antenna gains of both nodes. Despite the precision of the MLPL model, the propagation loss is estimated considering the distance between both nodes, regardless of their absolute and relative positions. This can lead to inaccurate estimations in complex environments with multi-path phenomena, where the propagation loss between the nodes can vary significantly for different relative positions, even though the distance is the same. Moreover, since both directions are treated equally, the MLPL model is unable to model asymmetric wireless channels, such as air-to-ground / ground-to-air wireless links \cite{khawaja2019survey}.


This paper proposes the \textbf{position-based ML Propagation Loss (P-MLPL) model} for ns-3. The P-MLPL model is built upon the MLPL model, thus inheriting its advantages over the trace-based simulation approach, such as the flexibility of allowing any network topology, node mobility, offered traffic and simulation duration. Based on network traces collected in an experimental testbed, the P-MLPL model estimates the propagation loss suffered by packets exchanged between a transmitter and a receiver, considering the absolute node's positions and the traffic direction. This improves the model's precision, especially in complex environments, while also enabling the modeling of asymmetric wireless channels. The P-MLPL model is composed of the path loss and the fast-fading components, which are trained with the network traces collected in the experimental testbed. The total propagation loss estimated by the P-MLPL model results from the sum of both components. Therefore, the P-MLPL model enables the development of fast and more precise digital twins of wireless networks in ns-3. The created digital twins allow the validation of novel solutions and the evaluation of their performance in realistic conditions, as long as the collected network traces can be used to characterize the dynamic conditions of the wireless channel (physical twin). Furthermore, it allows experimental testbeds to be digitally scaled in simulation, allowing the use of larger and more complex network topologies in realistic conditions.

To improve the computational complexity of ns-3 simulations and mitigate the overhead due to the use of ML algorithms, an internal \textbf{cache is added to the P-MLPL model} to save the latest ML values calculated, thus avoiding repeated calculations. This cache provides a speedup in the computational performance of the P-MLPL model, thereby minimizing the duration of corresponding ns-3 simulations. This optimization is especially important, since the P-MLPL model needs to calculate the estimated propagation loss for every packet generated in ns-3, which can rapidly scale with the number of nodes and offered traffic.

The P-MLPL model is validated with a test suite developed for the ns-3 module. Additionally, we evaluate the precision and performance of the model in a specific physical environment, using the respective experimental network traces. Specifically, we evaluate the precision of the path loss and the fast-fading models, by comparing their estimates with the real experimental values. Also, we evaluate the precision of the throughput values measured in ns-3 when using the P-MLPL model, by comparing them with the corresponding real values, as well as with existing propagation loss models available in ns-3. Finally, we evaluate the computational performance of the ns-3 simulations.


The rest of this paper is organized as follows.
\cref{MLPL-Section} explains the P-MLPL model.
\cref{Precision-Section} evaluates the P-MLPL model's estimation precision.
\cref{Performance-Section} analyzes the P-MLPL model's performance in ns-3 simulations.
\cref{Conclusions-Section} draws the conclusions and points out the future work.

\section{Position-Based ML Propagation Loss Model} \label{MLPL-Section}

This section presents the P-MLPL model, explaining its internal architecture, the structure of the ns-3 module and the dataset used to train the model. The P-MLPL model code is available in \cite{mlplGitlabRepo}.

\subsection{P-MLPL Model Architecture}

\begin{figure*}
    \centering
    \includegraphics[width=0.9\linewidth]{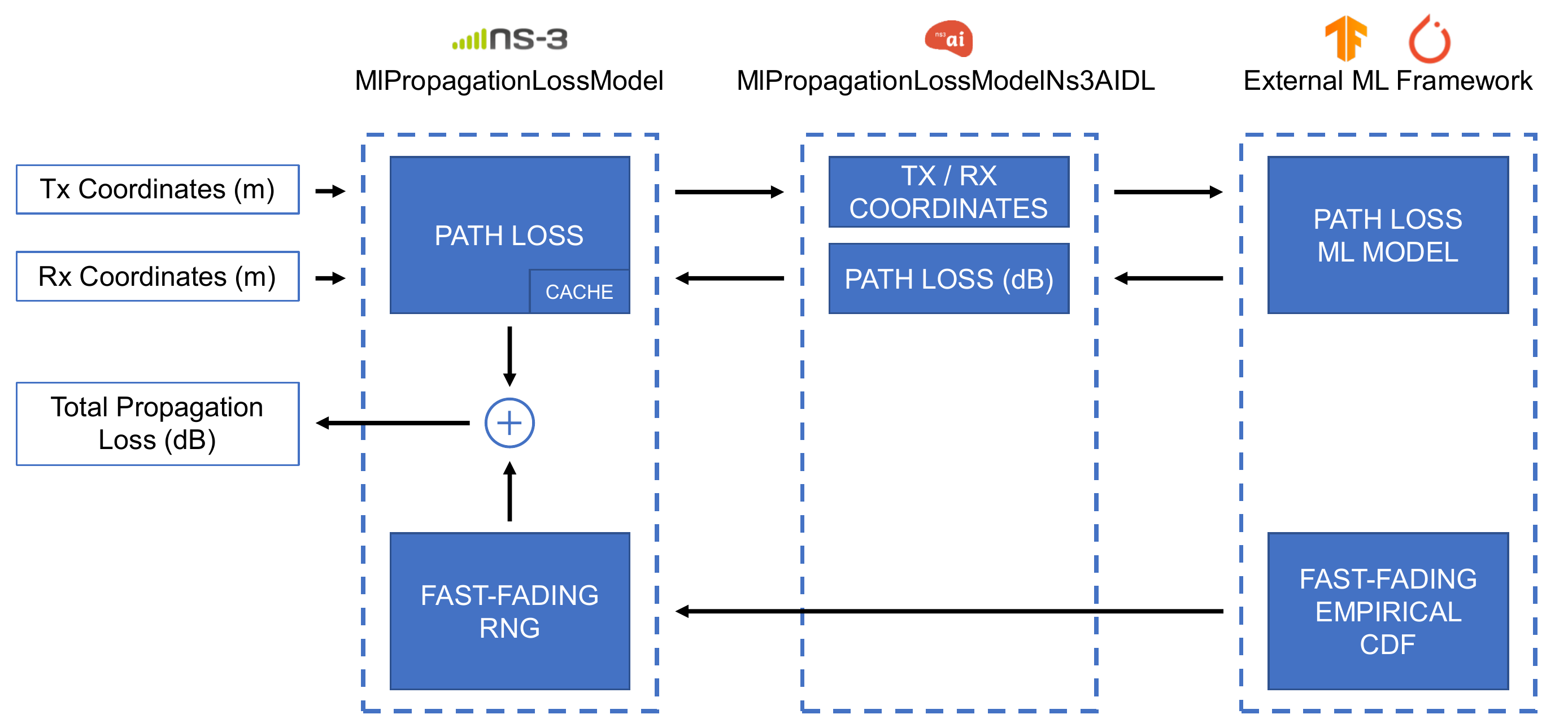}
    \caption{Position-Based MLPL Model Architecture, Depicting the Proposed MlPropagationLossModel, and the Interactions with MlPropagationLossModelNs3AIDL (ns3-ai) and External ML Frameworks}
    \label{MLPL-Figure: MLPL architecture}
\end{figure*}

The architecture of the P-MLPL model is shown in \cref{MLPL-Figure: MLPL architecture}. The P-MLPL ns-3 model is implemented in the \texttt{MlPropagationLossModel} class. This class is designed similarly to other propagation loss models in ns-3, by inheriting from the \texttt{PropagationLossModel} base class and overriding the \texttt{DoCalcRxPower()} virtual method. Internally, it is composed of the deterministic path loss ML model and the stochastic fast-fading model.

The path loss ML model uses a supervised learning algorithm to estimate the deterministic path loss of the wireless channel between a transmitter and a receiver. We use the ns3-ai module \cite{yin2020ns3}, available in the ns-3 App Store, to allow the use of external ML models developed with existing ML frameworks -- for instance, Tensorflow, PyTorch or SciPy -- without integrating them directly in ns-3. This module enables the exchange of data between ns-3 and an external computer process -- in this case, the external ML model -- via shared memory. We implement the \texttt{MlPropagationLossModelNs3AIDL} class to manage this interface, which inherits from the \texttt{Ns3AIDL} base class available in ns3-ai. To improve the computational complexity of the ns-3 simulation and mitigate the overhead of the ns3-ai module, an internal cache is added to the P-MLPL model to save the latest path loss ML queries. Whenever the P-MLPL model requires a new path loss value for a given pair of node positions, it first searches the value in its cache. If the value is found, it uses it immediately; otherwise, the P-MLPL model queries the path loss ML model for that value and saves it in the cache for future use. This mechanism avoids repeated queries to the ML model through the ns3-ai's shared memory, which are more expensive computational operations than retrieving values from memory.

The fast-fading component is modeled as a stochastic ergodic process that generates pseudo-random samples according to the empirical Cumulative Distribution Function (CDF) that fits the dataset samples representing the phenomena. The pseudo-random fast-fading samples are generated by a Random Number Generator (RNG) according to the fitted empirical CDF. To ensure repeatable and reproducible simulations, the fast-fading RNG is configured in ns-3, rather than in the external ML framework. This allows the RNG to be controlled by the same pseudo-random seed used in the ns-3 simulation, thus ensuring that the stream of pseudo-random samples are repeatable and consistent with the other RNGs in ns-3. Additionally, this design avoids the computational overhead associated with the shared memory mechanism of the ns3-ai module.

The total propagation loss for a given pair of node positions is calculated as the sum of the path loss for those positions and a pseudo-random sample from the fast-fading empirical CDF. When ns-3 requests the P-MLPL model to calculate the received power for a given packet sent by a transmitter to a receiver, the P-MLPL model returns the transmission power plus the calculated loss (path loss plus fast-fading). This value is later added to the antenna gains by ns-3, depending on the wireless models configured in the simulation.

\subsection{P-MLPL Module Structure}

The P-MLPL module for ns-3 contains the P-MLPL propagation loss model, as well as a set of helper scripts to train the ML models and manage the interface with external ML frameworks via the ns3-ai module.

Specifically, two helper Python scripts are provided to the users of the model. The \texttt{train\_ml\_propagation\_loss\_model.py} Python script implements the training of the path loss ML model and the fast-fading empirical CDF, exporting the corresponding models as files that are read by the ns-3 simulation. This script works offline without requiring parallel ns-3 simulations and must be run once before attempting to start ns-3 simulations, so that the ML models are available to be used by P-MLPL.

The \texttt{run\_ml\_propagation\_loss\_model.py} Python script initializes the \texttt{MlPropagationLossModelNs3AIDL} model, which manages the shared memory that enables the exchange of data between the \texttt{MlPropagationLossModel} and the external ML models.

Additionally, the P-MLPL ns-3 module contains the developed test suite that validates the architecture and functionality of the P-MLPL model. There is also an ns-3 simulation example that shows how to use the P-MLPL model, which was used to generate the results in \cref{Performance-Section}.

\subsection{Deterministic Path Loss ML Model} \label{MLPL-Section: Path loss}

The path loss component calculates the signal's propagation loss based on the distance between the transmitter and the receiver. It is based on a supervised learning model that estimates the deterministic path loss, in dB, for a given pair of transmitter and receiver positions.

We considered two supervised learning algorithms to train the path loss ML model: the eXtreme Gradient Boosting (XGBoost) algorithm and the Support Vector Regression (SVR) algorithm. These learning algorithms were selected due to the good results obtained when used with the MLPL model in \cite{almeida2022machine}. The XGBoost algorithm was implemented using its own Python library, whereas the SVR algorithm was implemented with the SciPy library. The path loss ML model is trained with the path loss samples in the dataset.

\subsection{Stochastic Fast-Fading Model} \label{MLPL-Section: Fast-fading}

The fast-fading component is modeled as a stochastic random variable characterized by the empirical CDF that best fits the fast-fading samples in the dataset. After the fitting process, the empirical CDF is exported to a CSV file as a collection of $ (X, Y) $ pairs, where $ X $ corresponds to the fast-fading loss, in dB, and $ Y $ corresponds to the respective CDF percentile rank in $ [ 0\%, 100\% ] $.

In order to generate a stream of pseudo-random samples according to the fast-fading empirical CDF, an \texttt{EmpiricalRandomVariable} RNG is used by the P-MLPL model. When the P-MLPL model is created, the RNG is configured with the empirical CDF's $ (X, Y) $ pairs contained in the exported CSV file. Whenever a fast-fading sample is requested from the P-MLPL model, a new pseudo-random value is sampled from the RNG.

\subsection{P-MLPL Propagation Loss Dataset} \label{MLPL-Section: Dataset}

The training of the P-MLPL model requires a dataset of propagation loss samples collected in an experimental testbed. The description of the dataset and the processing of its values are explained in the following sections.

\subsubsection{Dataset Format}

The P-MLPL dataset consists of a CSV file containing samples of the experimental propagation loss and the respective transmitter and receiver node positions.

The node positions are written in Cartesian coordinates $ (x, y, z) $, in meters, relative to a given origin point $ (0, 0, 0) $. The origin point can be arbitrarily selected by the dataset creators, provided that all coordinates in the dataset are consistent with this reference. Moreover, ns-3 simulations using the P-MLPL model must use the same origin point in their mobility models, so that the propagation loss estimations are precise.

The propagation loss values can be provided in two formats. In the first format, the propagation loss values, in dB, are provided directly in the dataset. In the second format, the propagation loss is calculated indirectly using other metrics provided in the dataset, including the received signal power, or the SNR, along with the noise floor. To calculate the propagation loss, an additional JSON file must be provided, containing information about the wireless network used in the experimental testbed, such as the Wi-Fi standard, the transmission power, the antenna gains, the channel frequency and the channel bandwidth.

Apart from these mandatory fields, more information can be optionally added to the samples, including the instantaneous throughput measured by the receiver node. Although these fields do not influence the training of the P-MLPL model, they provide additional information that can be used to calculate other performance metrics related to the model. One example is the precision of the throughput measured in ns-3 simulations; this is analyzed in \cref{Performance-Section: Throughput precision}.

\subsubsection{Dataset Outliers}

To improve the training of the ML models, the outlier points in the dataset are first removed. For each group of samples of the same node positions, the standard z-score $ { z = (x - \mu) / \sigma } $ is calculated for each sample $ x $, where $ \mu $ is the population mean and $ \sigma $ is the population standard deviation. We consider that a point is an outlier if its absolute z-score $ |z| $ is greater than 5.

\subsubsection{Decomposition of Propagation Loss into Path Loss and Fast-fading}

Since the dataset only contains the total propagation loss value, it is necessary to decompose it into the path loss and the fast-fading components. We assume that the fast-fading values follow a statistical distribution whose mean is 0 dB. This assumption enables the decomposition of the total propagation loss value into a deterministic path loss value plus a stochastic fast-fading random variable.

Thus, the path loss for each pair of node positions corresponds to the mean propagation loss value calculated using all samples of that pair of positions. The fast-fading values are calculated as the difference between the total propagation loss values and the path loss calculated for the corresponding node positions. The isolated path loss and fast-fading values are then used to train the corresponding ML models.

\subsection{P-MLPL Module Test Suite}

In order to validate the architecture and functionality of the P-MLPL ns-3 module, we developed a new test suite named \linebreak \texttt{MlPropagationLossModelTest}. The objectives of the test suite are two-fold. On the one hand, the test suite validates the exchange of data between the P-MLPL model and the external ML framework via the shared memory mechanism provided by the ns3-ai module, which is managed by the \texttt{run\_ml\_propagation\_loss\_model.py} helper script. On the other hand, the test suite validates the correctness of the propagation loss values calculated by the P-MLPL model, which is the sum of the path loss ML model predictions and the samples drawn from the fitted fast-fading empirical distribution.

To that end, this test suite contains a set of \emph{(node positions, propagation loss)} pairs taken from the example dataset contained in the ns-3 module. For each node position, the total propagation loss value calculated by the P-MLPL model is compared with the corresponding experimental value.

\section{P-MLPL Model Precision} \label{Precision-Section}

The precision of the P-MLPL model is evaluated in this section, including the precision of the path loss and the fast-fading models, as well as the overall overall propagation loss.

\subsection{Experimental Dataset and Evaluation Methodology} \label{Precision-Section: Dataset and evaluation}

In order to train the P-MLPL model and evaluate its precision, we created an experimental propagation loss dataset. This dataset was adapted from the larger dataset collected in the SIMBED project~\cite{simbedDataset} (SubExp3 -- run "08022019\_11.04.35"), which consists of experimental wireless network traces collected using Fed4FIRE+ testbeds \cite{fed4firePlus} in a warehouse environment. The experimental dataset used in this paper is available on the P-MLPL's GitLab repository \cite{mlplGitlabRepo}, with the name "position-dataset-example".

\begin{figure}
    \centering
    \includegraphics[width=0.8\linewidth]{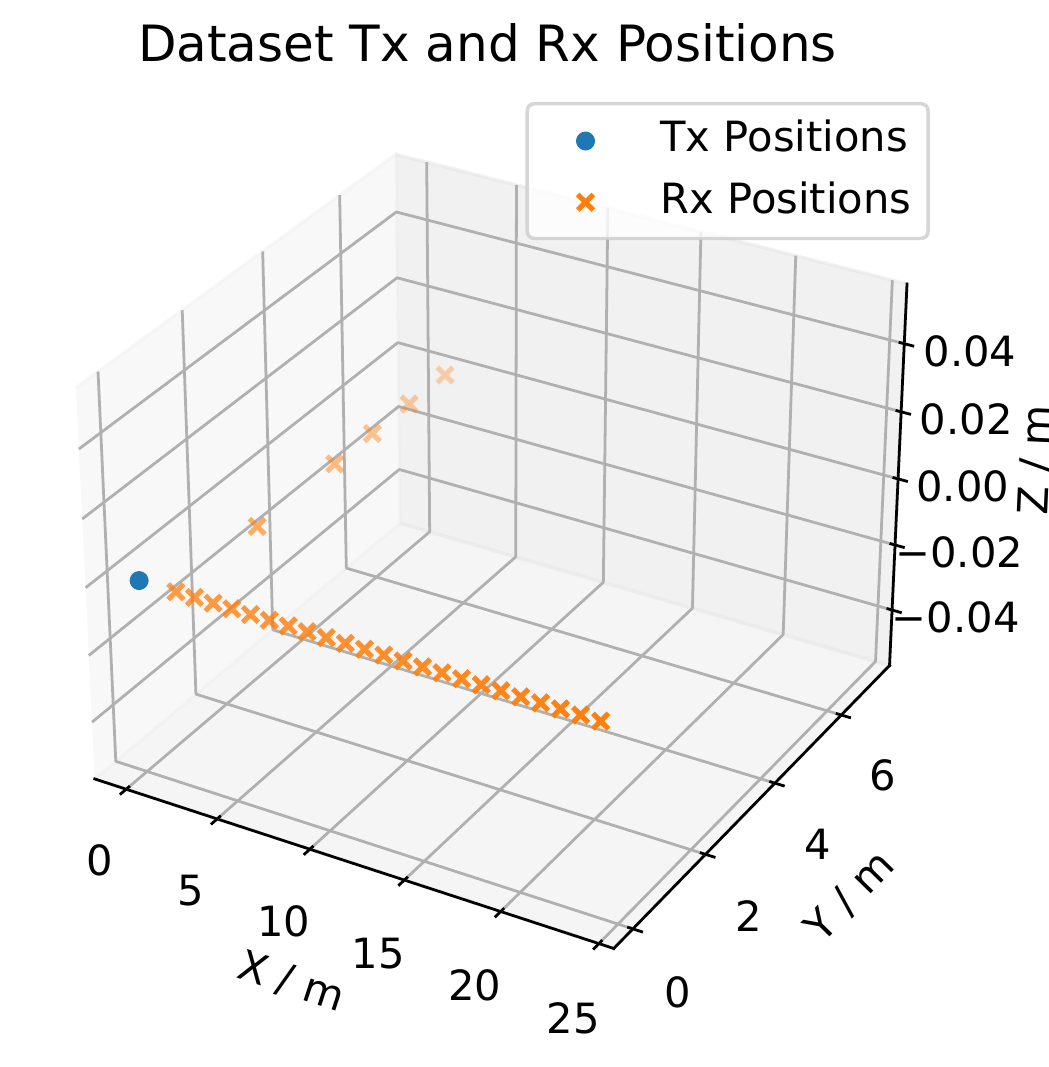}
    \caption{Set of Transmitter (Tx) and Receiver (Rx) Node Positions in the Experimental Dataset}
    \label{Precision-Figure: Dataset positions}
\end{figure}

The set of positions of the transmitter and receiver nodes contained in the experimental dataset is shown in \cref{Precision-Figure: Dataset positions}. The transmitter node is fixed at position $ (x, y, z) = (0, 0, 0) $, whereas the receiver node is moving away and towards the transmitter node in a straight line in the X-axis and the Y-axis. For each pair of \emph{(transmitter position, receiver position)}, the dataset contains the SNR, noise floor and throughput measured by the receiver node. The dataset values were collected considering the network parameters in \cref{Precision-Table: Dataset parameters}. It is worth noting that we used an effective antenna gain of -7 dBi for the antennas in both nodes. The gain of the antennas is a negative value since signal attenuators of 10 dB were used in-line with the 3 dBi antennas, to limit the signal's range in the warehouse.

\begin{table}
    \caption{Network Parameters of the Experimental Dataset}
    \label{Precision-Table: Dataset parameters}
    \begin{tabular}{c c}
        \toprule
        \textbf{Parameter} & \textbf{Value} \\
        \midrule
        Wi-Fi standard & IEEE 802.11a \\
        Wi-Fi MAC & Ad Hoc mode \\
        Tx power & 1 dBm \\
        Antenna gains & -7 dBi (3 dBi gain - 10 dBi attenuator) \\
        Channel frequency & 5220 MHz \\
        Channel bandwidth & 20 MHz \\
        MAC rate adaptation & Minstrel \\
        Application traffic & 54 Mbit/s UDP constant bitrate \\
        Packet size & 1400 bytes \\
        \bottomrule
    \end{tabular}
\end{table}

As explained in \cref{MLPL-Section: Dataset}, the dataset was pre-processed to remove the outliers, calculate the total propagation loss based on the SNR and noise values, and decompose the propagation loss into the path loss and fast-fading components. Then, the dataset was split in two sets: 1) the training set, containing 80\% of randomly-selected samples; and 2) the test set, with the remaining 20\% of samples.

\subsection{Path Loss ML Model Precision} \label{Precision-Section: Path loss precision}

The path loss ML model was trained with the path loss values in the training set. As referred in \cref{MLPL-Section: Path loss}, two supervised learning algorithms were considered for the path loss ML model: the eXtreme Gradient Boosting (XGBoost) and the Support Vector Regression (SVR).

The precision of the path loss ML model was evaluated using the path loss values in the test set. The real path loss values $ P_i $ of the test set were compared to the respective predictions $ \hat{P}_i $ by the ML model for the same set of node positions. The precision of the ML model was evaluated as the Mean Squared Error (MSE) calculated with all $ N $ samples of the test set, which is given by \cref{Precision-Equation: MSE}.

\begin{equation} \label{Precision-Equation: MSE}
    \textrm{MSE} = \frac{1}{N} \sum_{i=1}^{N} \left( P_i - \hat{P}_i \right) ^2
\end{equation}

\cref{Precision-Table: Path loss precision} presents the MSE for the path loss ML models trained with the XGBoost and the SVR learning algorithms. The results show that the \textbf{SVR model achieved an $ \textrm{MSE} = \SI{1.6}{dB}^2 $, resulting in very precise predictions of the path loss} for the set of positions in the test set. The XGBoost model achieved an $ \textrm{MSE} = \SI{4.3}{dB}^2 $, resulting in precise predictions of the path loss, although less precise than the SVR.

\begin{table}
    \caption{Path Loss ML Model Precision for Different ML Algorithms}
    \label{Precision-Table: Path loss precision}
    \begin{tabular}{c c}
        \toprule
        \textbf{ML Algorithm} & \textbf{Mean Squared Error (MSE)} \\
        \midrule
        XGBoost & $ \SI{4.3}{dB}^2 $ \\
        SVR & $ \SI{1.6}{dB}^2 $ \\
        \bottomrule
    \end{tabular}
\end{table}

\subsection{Fast-Fading Distribution Precision} \label{Precision-Section: Fast-fading precision}

As indicated in \cref{MLPL-Section: Fast-fading}, the fast-fading empirical CDF was fitted with the fast-fading values in the dataset. The precision of the fitted distribution was determined by comparing the CDF values of the fast-fading empirical CDF and the histogram of the original data. In this sense, we divided them into 30 bins and calculated the MSE between the CDF values of the fitted empirical CDF and the original data histogram.

\begin{figure}
    \centering
    \includegraphics[width=\linewidth]{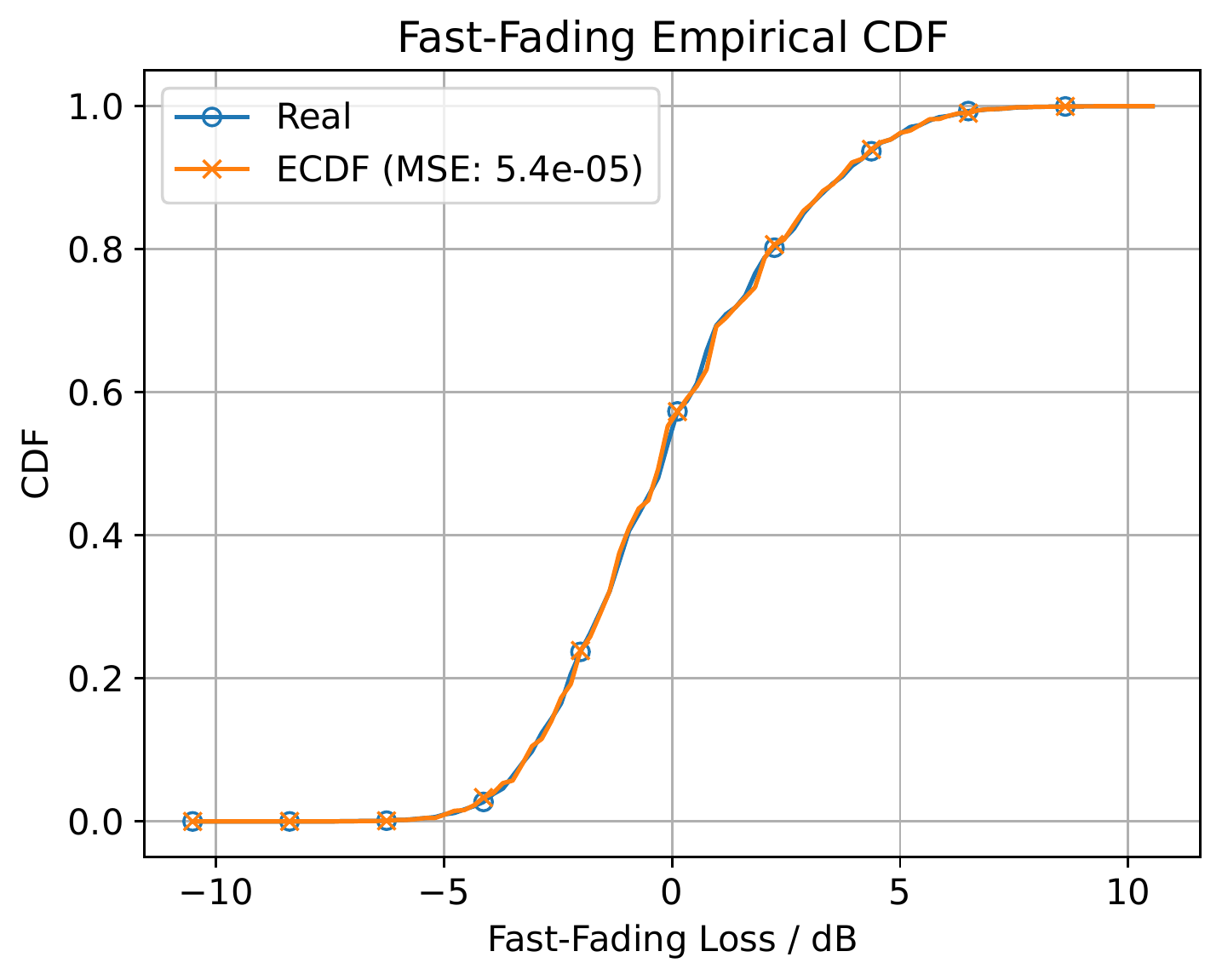}
    \caption{Fast-Fading Empirical CDF, Showing the Fitted Distribution, the Calculated MSE and the Original Data}
    \label{Precision-Figure: Fast-fading ECDF}
\end{figure}

The results of the fast-fading empirical CDF fitting are presented in \cref{Precision-Figure: Fast-fading ECDF}, showing the fast-fading empirical CDF, the real fast-fading values in the dataset, and the calculated MSE. It can be seen that the \textbf{fast-fading empirical CDF was perfectly fitted to the original data}, achieving an $ \textrm{MSE} = 5.4 \times 10^{-5} $. Thus, the empirical CDF provides a perfect representation of the fast-fading values in the dataset.

\subsection{P-MLPL Model Precision}

\begin{figure*}
    \centering
    \subfigure[P-MLPL (XGBoost) Prediction Error] {
        \includegraphics[width=0.45\linewidth]{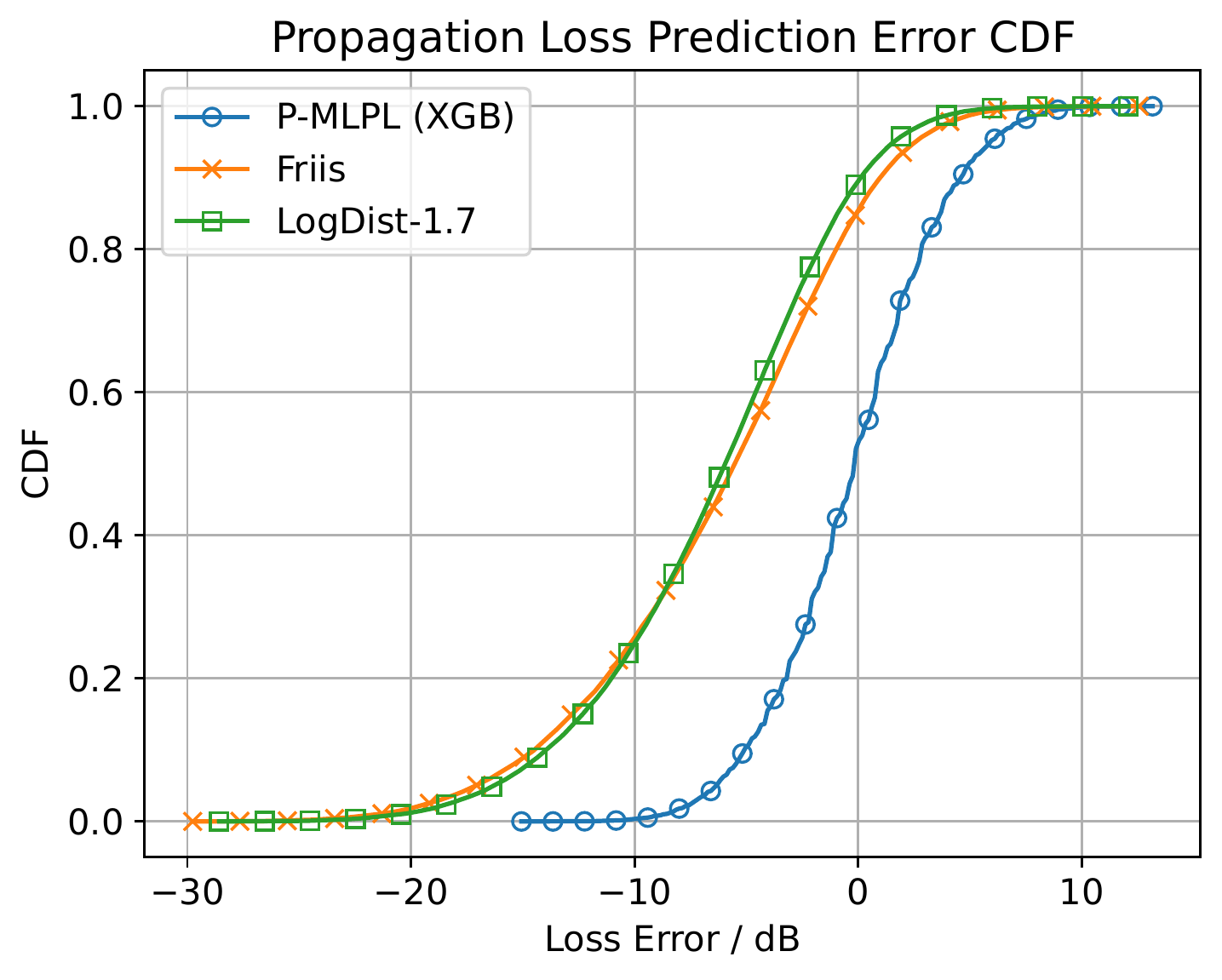}
    }
    \hfil
    \subfigure[P-MLPL (SVR) Prediction Error] {
        \includegraphics[width=0.45\linewidth]{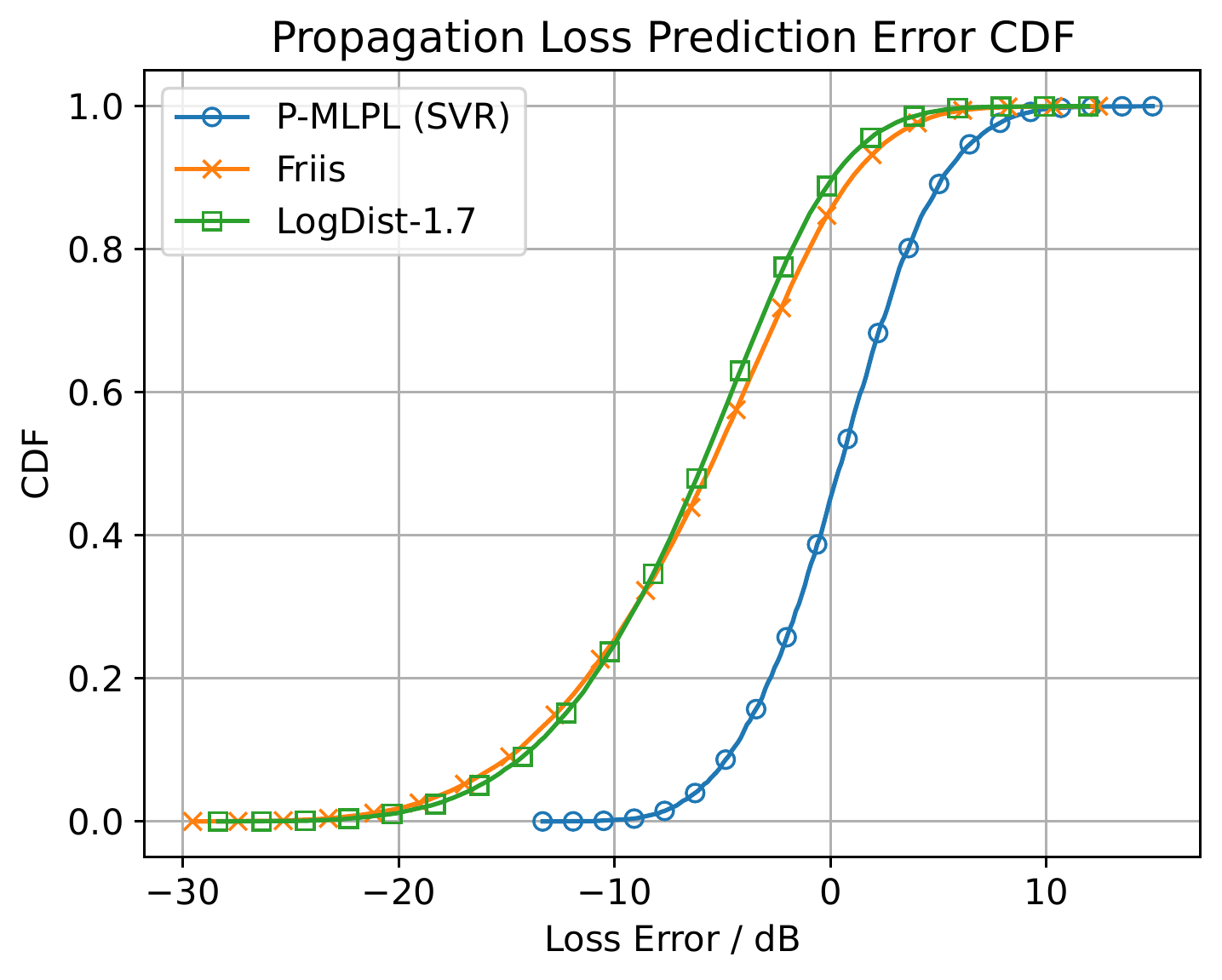}
    }
    \vfil
    \subfigure[P-MLPL (XGBoost) Absolute Prediction Error] {
        \includegraphics[width=0.45\linewidth]{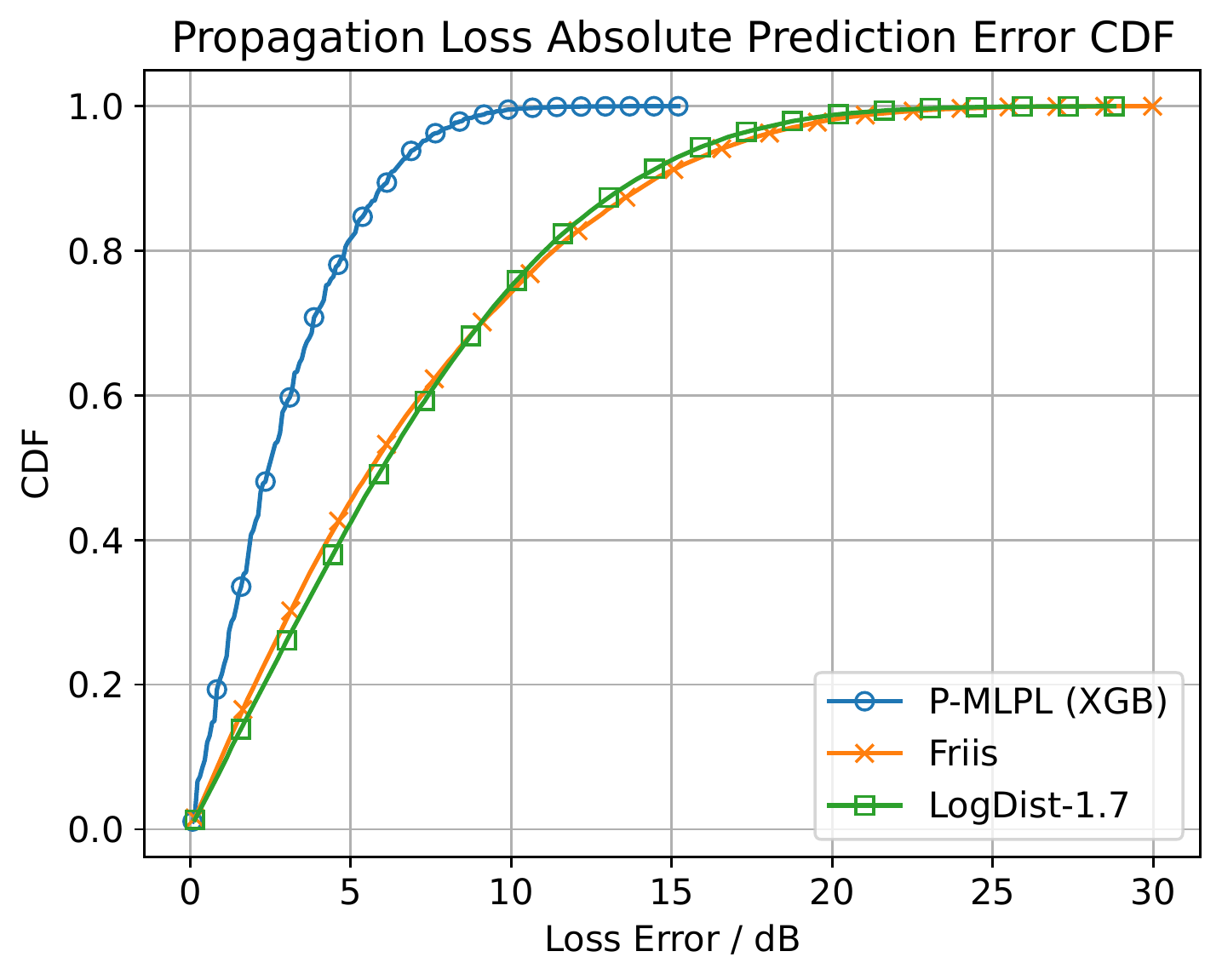}
    }
    \hfil
    \subfigure[P-MLPL (SVR) Absolute Prediction Error] {
        \includegraphics[width=0.45\linewidth]{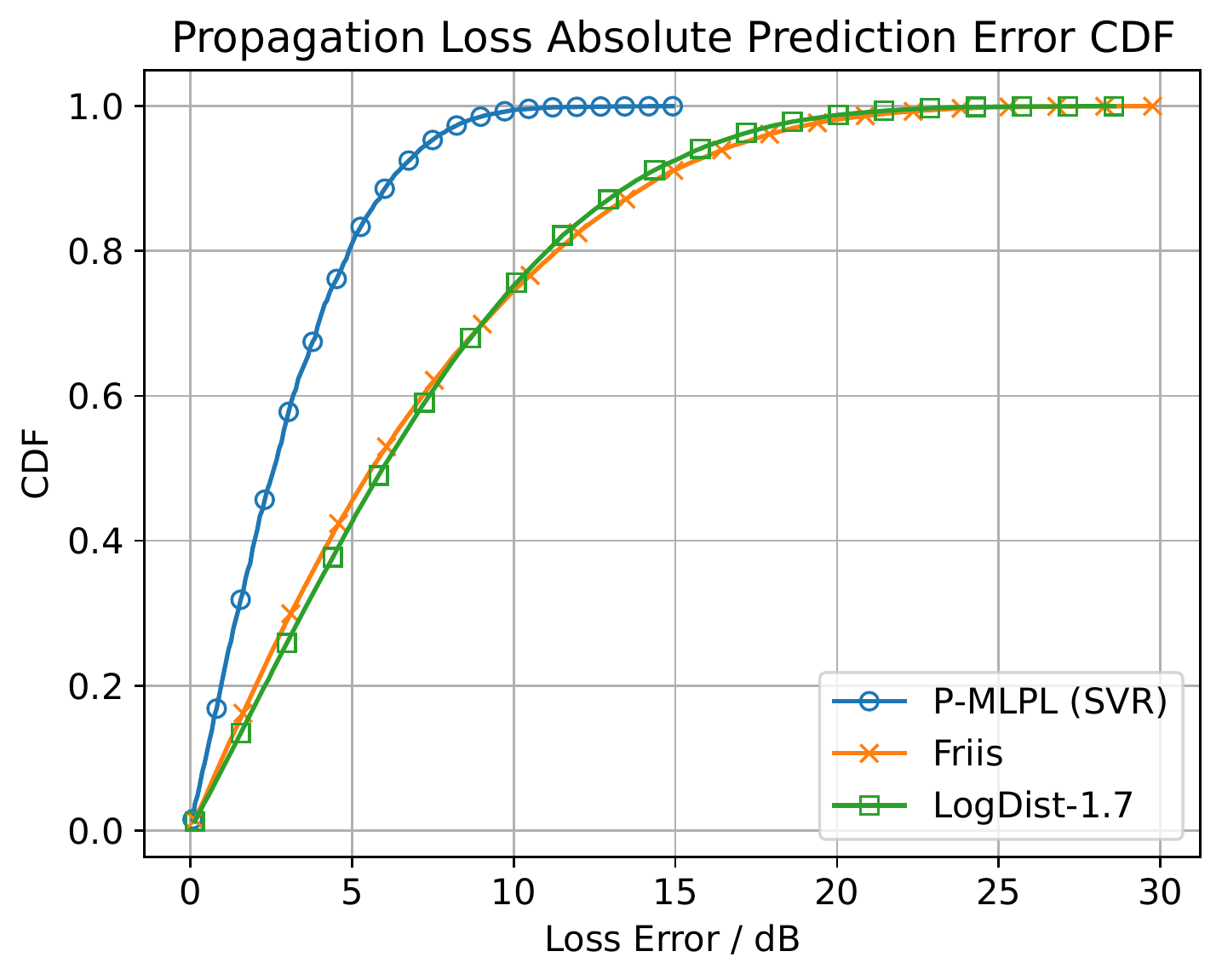}
    }
    \caption{CDF of the Real and Absolute Propagation Loss Prediction Errors by the P-MLPL Model, Compared to the Friis and the Log-Distance Models}
    \label{Precision-Figure: Propagation loss prediction error CDFs}
\end{figure*}

In addition to the individual evaluation of the path loss and fast-fading models, we also evaluated the precision of the full P-MLPL model. In this sense, we considered the two path loss ML models based on the XGBoost and SVR learning algorithms, which were combined with the best-fitted fast-fading distribution determined in \cref{Precision-Section: Fast-fading precision}.

The precision of the P-MLPL model was evaluated by comparing the propagation loss values in the test set to the corresponding estimations by the P-MLPL model. For each propagation loss value $ L_i $ in the test set, we calculated the prediction error of the P-MLPL model $ E^L_i = \hat{L}_i - L_i $ as the difference between the propagation loss predicted by the P-MLPL model $ \hat{L}_i $ and $ L_i $. Additionally, we calculated the prediction errors obtained with other propagation loss models available in ns-3, such as Friis and Log-Distance. We used the Log-Distance with path loss exponent $ \gamma = 1.7 $, which provided the closest match to the original data. We also added a Normal fast-fading distribution to both the Friis and Log-Distance models, with a mean $ \mu = \SI{0}{dB} $ and a standard deviation $ \sigma = \SI{3}{dB} $ that matches the distribution of fast-fading values in the dataset.

The CDF of the real and absolute prediction errors are shown in \cref{Precision-Figure: Propagation loss prediction error CDFs}. The results demonstrate that \textbf{the P-MLPL model is the most precise propagation loss model}, both when using the XGBoost and the SVR ML algorithms. In fact, analyzing the median value of the absolute error CDF, the P-MLPL model was able to predict the total propagation loss between the transmitter and the receiver with an error up to 0.5x than the baseline models. Moreover, the results show that the precision of the SVR path loss model is similar to the XGBoost model, with a median absolute error of 2.5 dB. In terms of the real prediction errors, the results reveal that half of the P-MLPL model predictions were either above or below the real propagation loss values, which matches the fitted fast-fading distribution. On the other hand, the Friis and Log-Distance baseline path loss models were too optimistic in this scenario, given that more than 90\% of the predictions were below the real losses.

\section{P-MLPL Model Performance in ns-3 Simulations} \label{Performance-Section}

This section analyzes the performance of the P-MLPL model in the context of ns-3 simulations, in terms of the precision of the throughput obtained in simulation and the computational performance of the model.

\subsection{ns-3 Simulation Set-Up and Parameters}

To analyze the P-MLPL model performance, we used the experimental dataset described in \cref{Precision-Section: Dataset and evaluation}. The set-up of the ns-3 simulation replicated the experimental environment where the dataset was collected, whose parameters are shown in \cref{Precision-Table: Dataset parameters}. We used the ns-3.37 version configured with the default parameters, except the ones referred herein. We set the minimum RSSI preamble detection threshold to -90 dBm, so that the lowest data rates could be used, replicating the behavior of the experimental wireless networks cards.

In order to evaluate the precision of the throughput obtained in simulation using the P-MLPL model, we replicated the positions of the transmitter and receiver nodes available in the dataset. For each unique pair of node positions, the transmitter sent a UDP constant bitrate traffic of 54 Mbit/s to the receiver during 5s, after an initialization period of 1s to stabilize the Minstrel rate adaptation algorithm and the MAC queues. The generated traffic matched the traffic that was generated when collecting the dataset, which was selected to ensure that the connection is always fully loaded, as the offered load is always above the link capacity. Then, the average throughput received by the receiver node was calculated and saved.

In addition to the P-MLPL model, we performed the same test for the Friis and Log-Distance models, to compare the precision of the throughput obtained with the P-MLPL model against these baselines.

\subsection{ns-3 Throughput Precision} \label{Performance-Section: Throughput precision}

After executing the tests explained in the previous section, we analyzed the precision of the throughput obtained in ns-3 simulations using the P-MLPL model. For each pair of node positions, we calculated the throughput error $ E^T_i = \hat{T_i} - T_i $ as the difference between the throughput measured in ns-3 $ \hat{T}_i $, and the respective real throughput $ T_i $ in the dataset.

\begin{figure}
    \centering
    \subfigure[Average Throughput Error]{
        \includegraphics[width=\linewidth]{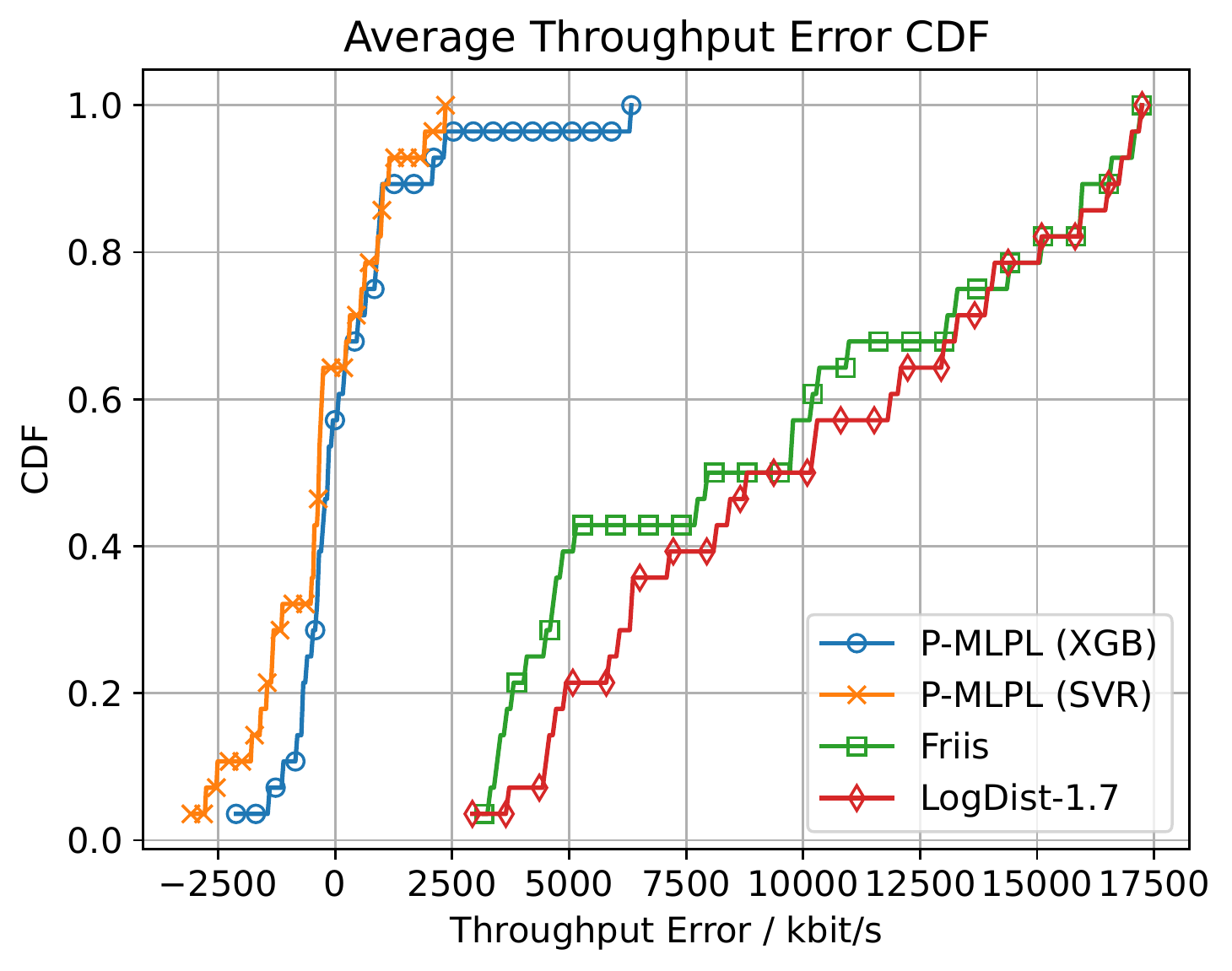}
    }
    \vfil
    \subfigure[Average Absolute Throughput Error]{
        \includegraphics[width=\linewidth]{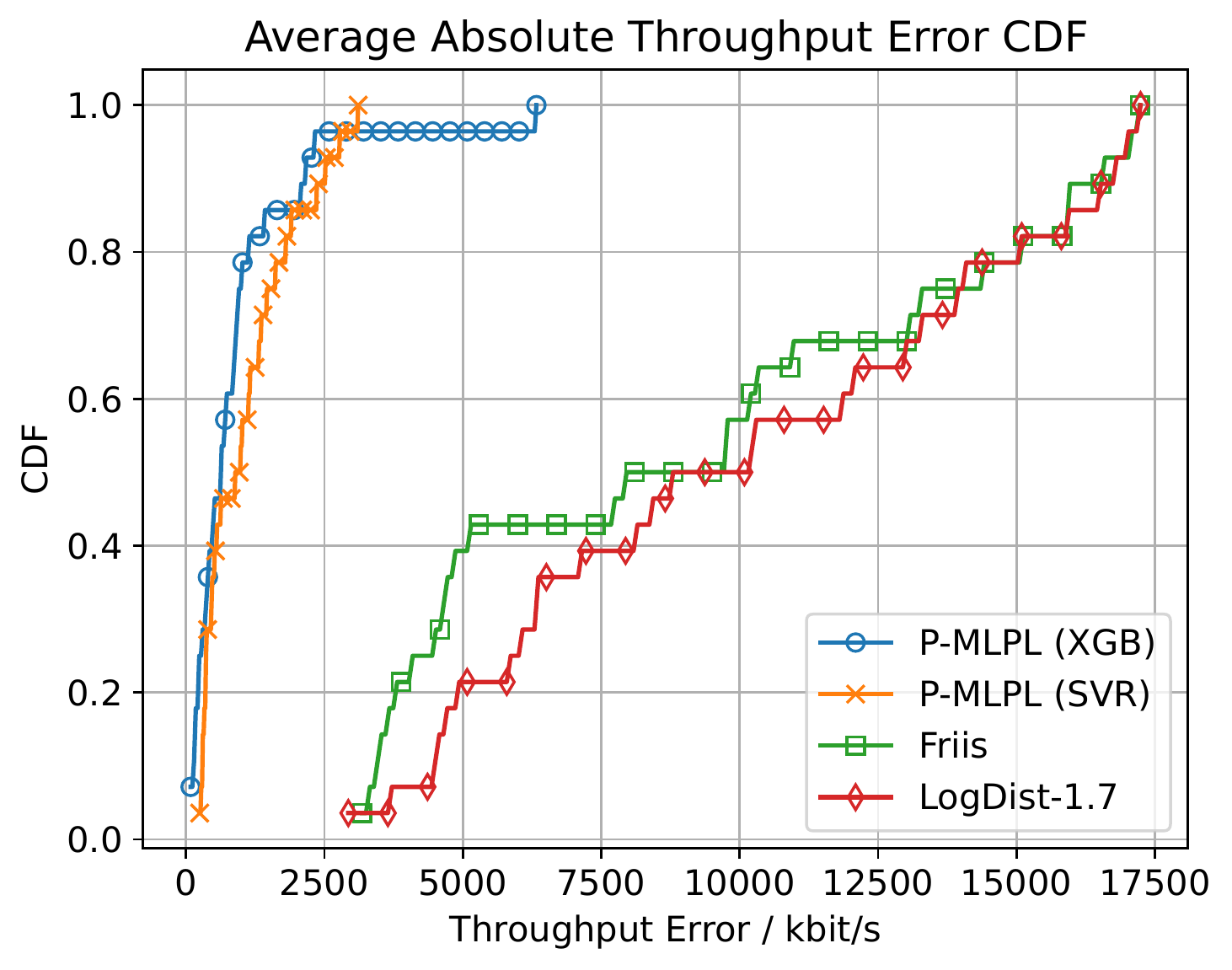}
    }
    \caption{CDF of the Average Throughput Error Compared to the Real Throughput Collected in the Experimental Dataset}
    \label{Performance-Figure: Throughput error CDF}
\end{figure}

The CDF of the real and absolute throughput errors are shown in \cref{Performance-Figure: Throughput error CDF}. The results demonstrate that the P-MLPL model was able to \textbf{accurately reproduce the experimental throughput} of the dataset. Analyzing the 90th percentile, both the P-MLPL models based on XGBoost and SVR predicted the throughput with an absolute error up to 2.5 Mbit/s. Although the XGBoost provided better throughput precision than the SVR model, both ML models provided higher precision, compared to the the Friis and Log-Distance models, which had absolute prediction errors up to 16 Mbit/s.

Moreover, these results confirm the conclusions drawn in \cref{Precision-Section}. The P-MLPL model's increased precision in estimating the propagation loss translated into an increased throughput estimation precision. The optimistic estimations of the propagation loss by the Friis and Log-Distance models translated into optimistic throughput estimations.

\subsection{P-MLPL Model Computational Performance}

In order to evaluate the computational performance of the P-MLPL model, with and without caching propagation loss ML queries, we analyzed the time needed to finish ns-3 simulation runs using the P-MLPL model. For this purpose, we analyzed the duration of the ns-3 simulations executed in \cref{Performance-Section: Throughput precision}. For each propagation loss model and cache configuration, we ran 10 ns-3 simulations runs using different seeds to calculate the 95\% confidence intervals. The ns-3 simulations were executed one-by-one in an Ubuntu 22.04 with 8 CPU cores and 16 GB of RAM. The ns-3.37 simulator was configured and built with the \emph{optimized} build profile.

The results of the tests can be seen in \cref{Performance-Table: ns-3 simulation duration}. The results show that, with the ML cache disabled, the P-MLPL model based on the SVR ML model was the fastest ML model, finishing a simulation run in 1m 16s on average. On the other hand, the P-MLPL model based on XGBoost was 25x slower than SVR, finishing a simulation run in 31m 45s on average. These results demonstrate that the \textbf{P-MLPL model based on SVR is a much faster alternative to the P-MLPL model based on XGBoost}, with similar propagation loss and throughput precision, making it the best choice for most of the scenarios.

\begin{table}
    \caption{Mean ns-3 Simulation Duration and 95\% Confidence Interval for Different Propagation Loss Models and Cache Configurations}
    \label{Performance-Table: ns-3 simulation duration}
    \begin{tabular}{c c c}
        \toprule
        \textbf{Propagation Loss} & \textbf{Duration} & \textbf{Duration} \\
        \textbf{Model} & \textbf{(Without Cache)} & \textbf{(With Cache)} \\
        \midrule
        P-MLPL (SVR)     & 1m 16s $ \pm $ 3s   & 12s $ \pm $ 2s \\
        P-MLPL (XGBoost) & 31m 45s $ \pm $ 28s & 13s $ \pm $ 1s \\
        Friis            & 19s $ \pm $ 2s      & 19s $ \pm $ 2s \\
        Log-Distance     & 27s $ \pm $ 1s      & 27s $ \pm $ 1s \\
        \bottomrule
    \end{tabular}
\end{table}

With the ML cache enabled, the duration of the simulations reduced significantly. The P-MLPL model based on SVR reduced the duration of a simulation run to only 12s on average, representing a speedup of 6.3x. Similarly, the P-MLPL model based on XGBoost reduced the duration to 13s, representing a speedup of 147x. It is worth noting that the simulation set-up used in these tests benefits the caching technique, since the nodes adopt constant positions throughout each throughput monitoring period. Still, this condition is often used in many simulation scenarios, demonstrating the usefulness of this optimization.

To analyze the performance impact of the ML models in the overall ns-3 simulation duration, we also executed the same tests with the baseline propagation loss models Friis and Log-Distance. Since these propagation loss models do not use external ML models nor caching, they provide reference times for ns-3 simulations without the ML overhead and cache optimization. On average, when using the Friis propagation loss model, the ns-3 simulation run in 19s, whereas when using the Log-Distance model, it finished in 27s.

These results demonstrate that, although the ML overhead increases the simulation duration, \textbf{caching ML queries not only mitigates this overhead, but also significantly improves the performance of the simulation}, even surpassing the performance of existing propagation loss models in ns-3 that do not use ML. Furthermore, the results also show that, although the Friis and Log-Distance models do not use external ML frameworks, the impact of repetitive calculations for the same set of node positions deteriorates the overall performance of ns-3 simulations. The application of the caching optimization technique to existing models in ns-3 is worthy of being explored in the future.

\section{Conclusions and Future Work} \label{Conclusions-Section}

In this paper, we proposed the position-based ML Propagation Loss Model (P-MLPL) to enable the creation of fast and more precise digital twins of wireless networks in ns-3, which is essential to develop and validate next-generation wireless networks solutions. Based on network traces collected in an experimental testbed, the P-MLPL model is able to estimate the propagation loss suffered by packets exchanged between a transmitter and a receiver, considering their absolute positions and the traffic direction. The propagation loss is calculated as the sum of the deterministic path loss plus a stochastic fast-fading value.

The results show that, for the scenario considered in this paper, the P-MLPL model can predict the propagation loss with a median error of 2.5 dB, which corresponds to 0.5x the error of the Friis and Log-Distance models available in ns-3. The increased precision in estimating the propagation loss translates into an increased precision of the throughput measured in ns-3, with an error up to 2.5 Mbit/s, when compared to the real values collected in the experimental testbed. Moreover, both the XGBoost and the SVR learning algorithms used to train the path loss model demonstrated similar estimation precision.

The use of an internal cache in the P-MLPL model to save repeated ML queries represented a speedup of up to 147x in the duration of a simulation run. Despite the overhead of the ML queries via ns3-ai's shared memory, the computational performance of the cache-enabled P-MLPL model even surpassed the performance of existing propagation loss models in ns-3 that do not use ML. These results inspire the application of this optimization technique to the existing models in ns-3 to benefit from the same speedup.

As future work, we plan to submit the P-MLPL model to the ns-3 App Store. Additionally, we plan to explore the use of neural networks as an alternative to the XGBoost and SVR supervised learning algorithms.

\begin{acks}
This article is a result of the project "DECARBONIZE – DEvelopment of strategies and policies based on energy and non-energy applications towards CARBON neutrality via digitalization for citIZEns and society" (NORTE-01-0145-FEDER-000065), supported by Norte Portugal Regional Operational Programme (NORTE 2020), under the PORTUGAL 2020 Partnership Agreement, through the European Regional Development Fund (ERDF).
\end{acks}

\balance
\bibliographystyle{ACM-Reference-Format}
\bibliography{references}

\end{document}